# Probing the Hydrogen Enhanced Near-Field Emission of ITO without a Vacuum-Gap

*Jacob L. Poole[*,1], Yang Yu[1,2], and Paul R. Ohodnicki[1]*

Dr. J. L. Poole, Dr. Y. Yu, and Dr. P. R. Ohodnicki
[1]National Energy Technology Laboratory, 626 Cochrans Mill Rd, Pittsburgh, PA 15236, USA
[2]AECOM, 626 Cochrans Mill Rd, Pittsburgh, PA 15236, USA
E-mail: JacobLorenziPoole@gmail.com

Thermal fluctuations of charged particles, fluctuations akin to Brownian motion, can excite optical surface-states leading to the concept of the thermal near-field, which is a highly localized, and, therefore, evanescent optical density of states that exist at distances much less than the thermal emission wavelength. By tunneling the surface charge emitted photons into nearby waveguides, the thermally excitable near-field optical density of states can be enhanced, engineered, and efficiently extracted to the far-field for observation. With this technique, the plasmonic thermal near-field of a 10nm thick ITO film, known to have plasmonic activity in the 1500nm wavelength region, was probed under external illumination and by thermal excitation at 873K. The results confirm that waveguides provide a large density of optical channels with spatial overlap and k-vector matching to facilitate plasmon de-excitation in the near-field through photon tunneling for extraction into the far-field. Furthermore, it is shown that the thermal near-field can be observed without the introduction of a vacuum-gap, a feature unique to this particular method.

## 1. Introduction

The thermal near-field, capable of generating thermal radiation many orders of magnitude greater than the commonly known far-field, is an active area of research as it is a place where Plank's theory of blackbody radiation is known to break down.[1] All matter, at temperatures greater than absolute zero, contains thermal energy stored in lattice and electronic vibrational modes. Temperature differentials give rise to the flow of heat, which, in the absence of conductive and convective heat transfer, is in the form of electromagnetic radiation emitted by the thermal motion of charged particles. It is known that surface-plasmon polaritons (delocalized charge-density fluctuations), surface-phonon polaritons (localized charge density fluctuations in polar dielectrics), and surface adsorbate vibrational-modes, are mechanisms that can give rise to these enhanced near-field optical density of states.[2, 3] Metal films such as silver and gold are known to have surface-plasmon polariton resonances in the UV-vis wavelength range, whereas silicon carbide and silica are polar materials known to have surface-phonon polariton type resonances at much longer wavelengths.[2-4, 5] On the other hand, there are a number of doped metal oxides and metal nitrides that exhibit plasmonic resonances at optical frequencies.[6] Nanoparticles of plasmonic materials such as gold and silver can assist in scattering the thermal near-field to the far-field, due to the existence of localized surface plasmon resonance.[7]

The thermal near-field is challenging to observe as it is an optical density of states confined to distance much less than the thermal emission wavelength, and it is evanescent having k-vectors that are parallel to the surface. To date, complicated techniques such as SNOM and grating assisted coupling were utilized to probe this physics. SNOM consists of bringing coated AFM tips near surfaces to scatter the surface waves to the far-field for observation.[2, 5, 8] Whereas, resonant structures such as gratings scatter the near-field into the far field, with which the thermally excited surface plasmon resonance of gold and tungsten has been observed.[9] In a recent publication, we have demonstrated the capability of optical fiber for observing the thermal near-field of a TiO$_2$ film through tunneling, as waveguides inherently provide the necessary combination of spatial overlap of the probability density with the evanescent tail, as well as k-vector matching.[10]

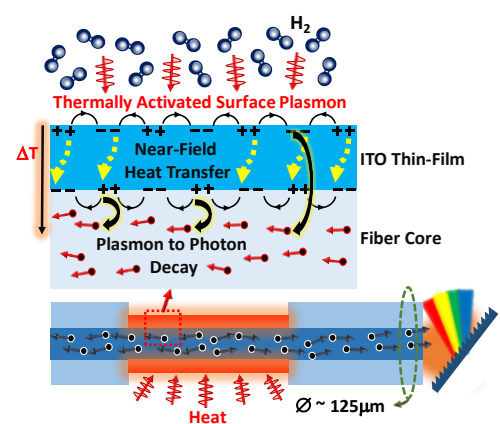

**Figure 1.** Schematic illustrating the thermal energy driven surface plasmon resonance of an indium tin oxide (ITO) thin-film, and the subsequent extraction of the thermal near-field radiation, generated by the de-excitation of the thermally driven charge density wave, through photon tunneling into an optical waveguide.

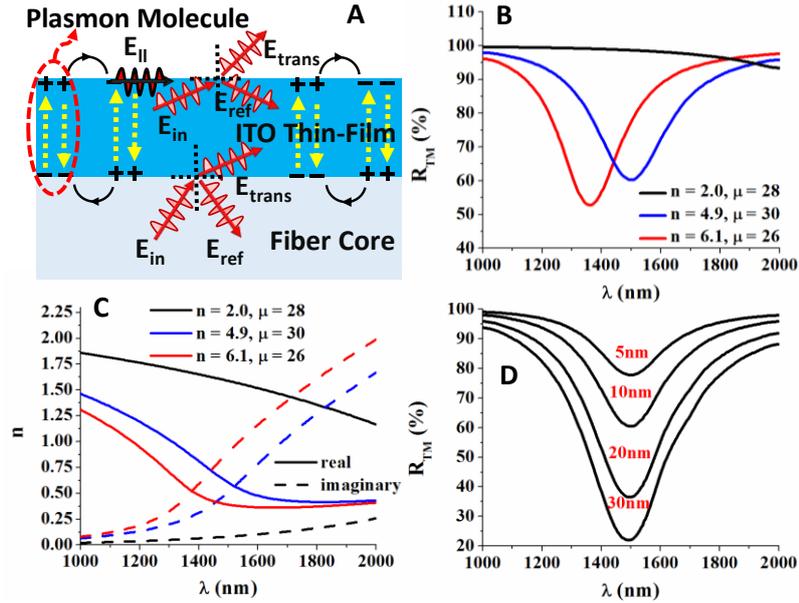

**Figure 2.** (A) Schematic illustrating the structure of the proposed ITO thin-film integrated waveguide. (B) The transverse magnetic reflectance ($R_{TM}$) obtained using the rigorous transfer matrix method for a film thickness of 10nm and various carrier concentrations (n in units of $10^{20}$ cm$^{-3}$) and mobility (μ in units of cm$^2$V$^{-1}$s$^{-1}$). (C) Real and imaginary parts of the refractive index associated with the carrier concentrations and mobility values. (D) Calculated variations in the transverse magnetic reflectance as a function of film thickness at around resonance.

In this work, we report experimental observation of the thermal-energy driven near-field optical density of states of a 10nm ITO thin-film at plasmon resonance, and the subsequent extraction to the far-field by tunneling into optical channels provided by a fiber-type waveguide (**Figure 1**). The presented results support the foundation of thermal near-field observation with waveguides, having important applications in thermal energy powered sensors and thermo-photonic energy harvesting, along with providing an inexpensive and practical near-field probing technique.[10, 11] The need for materials that are both high temperature stable and support surface plasmons resonances in the UV-vis and NIR wavelength ranges is clear, given the high value of the possible applications.

## 2. Waveguide Plasmon Interactions

In the constructed apparatus, the thickness of the film is much less than the thermal emission wavelength, being only 10nm thick. Thus, when the thin-film integrated optical waveguide is placed in a heated environment, the temperature differential that is developed facilitates the flow of heat from the source to the thin-film. The film then tunnels a spectrally narrow near-field surface plasmon-based emission into the optical fiber for extraction to a spectrometer. At the outer surface of the film the thermal energy excites a charge density wave which has several optical channels of de-excitation, one of which is the direct tunneling of photons into the waveguide. A surface charge density wave can be excited at the inner surface of the film, as well. Therefore, the outer hotter side of the film can transfer heat to the inner side of the film through tunneling and through dipole-dipole interactions, developing a communication channel between the outer and inner sides through near-field heat transfer, forming a coupled plasmonic molecular state.[3] Not unlike the bond formation of atoms making up molecules.

The conditions for the existence of an optically excitable surface charge density wave (plasmonic resonance) were examined using the rigorous transfer matrix method.[12] The transverse magnetic polarization (RTM) only has electric field components with surface-parallel k-vectors (Figure 1A) with the overall structure composed of a silica core, a 10nm ITO film, and an air interface. Therefore, only the RTM polarization shows a resonant dip associated with surface plasmon resonance (Figure 2B), examined at the critical angle inside of the waveguide structure, and located at the intersection of the real and imaginary components of the refractive index (Figure 2C), with nominal permittivity values obtained from Holman et al.[13] With an increase in the carrier concentration from 4.9 to $6.1 \times 10^{20} cm^{-3}$, the resonance peak of the transverse magnetic reflection shifts to lower wavelengths, when the Hall mobility is held relatively constant.

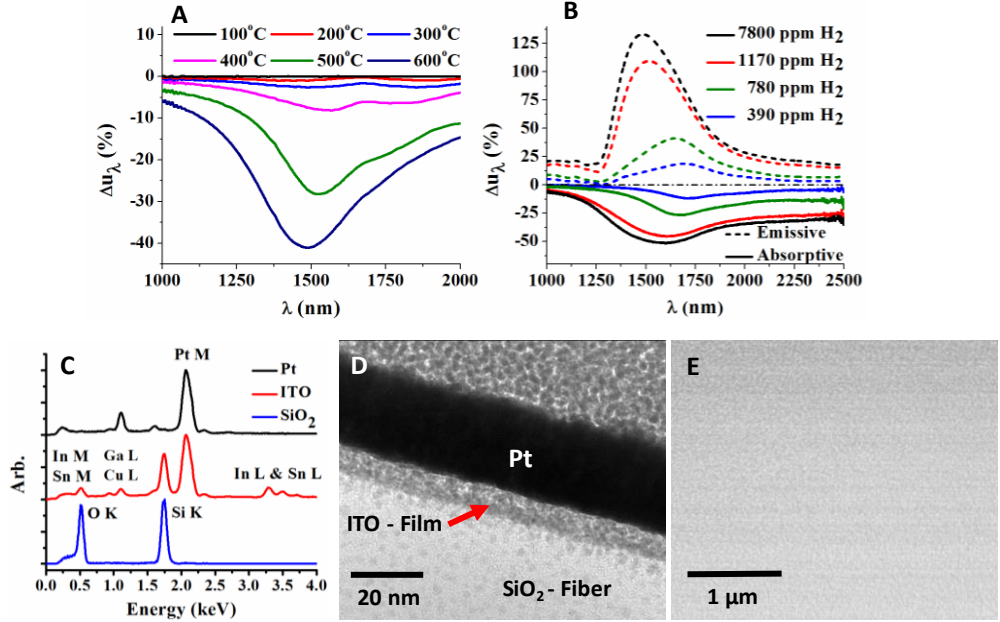

**Figure 3.** (A) Change in the relative spectra energy density $\Delta u_\lambda = (u_\lambda - u_{\lambda o})/u_{\lambda o}$ upon heating from room temperature to 873K in $N_2$. (B) Measured emissive and absorptive near-field optical density of states at plasmonic resonance of a 10nm ITO thin-film at 873K in various concentrations of $H_2$ balanced with $N_2$, normalized to 0ppm of $H_2$. The dashed lines correspond to thermal near-field emission based measurements, whereas the solid lines are obtained when probed with an external light source. (C) Cross sectional EDS spot analysis indicating the presence of In and Sn in the film. (D) Cross sectional TEM image showing the film thickness to be 10nm, obtained with platinum assisted FIB sectioning and lift-out. (D) BSE SEM image of the surface of a representative ITO film.

At a carrier concentration of $2 \times 10^{20} cm^{-3}$ the plasmon resonance has shifted to a much higher wavelength, outside of the current instrument range, indicating the sensitivity of resonance to this parameter. The dependence of the transverse magnetic reflectance on the film thickness is as shown in **Figure 2D**, where an increasing film thickness is accompanied by an increase of loss without altering the location of resonance, and should be explainable by an increased interaction-length.

### 3. Thermal Near-Field Measurement

Heating up the ITO film in an $N_2$ background, incrementally, provided the data in **Figure 3**, showing a resonance shift to lower wavelengths and an increase in the absorptivity, indicating that the free carrier concentration has increased, when referenced to room temperature. A clearly observable resonance peak is noted near 1500nm, indicating that the manufactured film has a sufficiently high carrier concentration and associated Hall mobility to sustain a surface charge density-wave, which, according to Figure 2B, is accomplished at 1550nm with a conductivity of approximately $2400 Scm^{-1}$. When probed with external illumination, successive changes are noted in the plasmon resonance upon exposure to varied concentrations of $H_2$ (**Figure 3B** - solid line). With an increasing $H_2$ partial pressure, a shift of the plasmon resonance to lower wavelengths and an increase in the absorbed light is observed, explained by an increase in the free electron concentration.

Given that it is challenging to measure absolute absorption in the current configuration, a relative change is measured upon referencing to 1% $O_2$ in $N_2$ background, instead, shown as $\Delta u_\lambda$ in % defined as $(u_\lambda - u_{\lambda o})/u_{\lambda o}$. Referencing to an oxidizing background atmosphere which is subsequently replaced with one containing an $H_2$ partial pressure, allowed the observation of free carrier concentration induced changes in the plasmon resonance both under external probing and when observing the near-field emission of the ITO surface-plasmon charge-density-wave. Upon removing the external illumination, the extracted thermal radiation was measured and is shown in **Figure 3B** (dashed line). The increase in the free carrier concentration will result in a more well defined surface plasmon wave, which is expected to be accompanied by an emission increase and a resonance shift to shorter wavelengths, as is observed. In comparing the magnitude of the relative spectral energy density of the thermal emission with the absorptive response, some discrepancies are noted. An asymmetry is inherent in the current realization from the perspective of heat flow, and filtering induced by the waveguide modifications due to thin-film integration which result in modal mismatches. However, it is not believed that Kirchoff's theory is violated, which states that absorptivity must equal emissivity, as it is more likely that the asymmetry inherent in the system is the source of the apparent violation. Using platinum assisted functional ion beam a section of the film from an ITO coated fiber was lifted and imaged with a transmission electron microscope yielding a film thickness of ~10nm (**Figure 3D**), along with an EDS spot scan spectral analysis confirming the presence of In and Sn in the region representative of the thin-film location (**Figure 3C**). It was found that exposing the ITO film to hydrogen concentrations above the explored range causes irreversible changes, an observation which we hope will motivate the exploration of plasmonic materials favorable to higher temperatures.

## 4. Conclusion

We report experimental evidence demonstrating the thermally excitable near-field optical density of states of a 10nm ITO thin-film at plasmonic resonance, without the need for a vacuum gap. The nearby waveguide provides a large density of optical channels with the necessary spatial overlap and k-vector matching to change the preferred mode of de-excitation to photon tunneling allowing for the extraction of the near-field emission into the far-field for observation. A configuration that boasts simplicity in comparison with the previous employed methods, such as SNOM where a tip is brought into near-contact with a surface to scatter the near-field into the far-field, or resonant scattering with gratings.[2, 5, 8, 9] The $H_2$ partial pressure induced modulation in the carrier concentration and Hall mobility has a clear effect on the emission spectrum, enhancing the resonant thermally emitted light of the ITO film by 125%. These observations hold great promise for the development of thermally powered sensors and novel thermal energy harvesting methods, having a geometrical footprint of a human hair, along with new instrumentation for probing the thermal near-field.

## 5. Methods

A 10nm thin-film of ITO was coated on the core of a 105μm core optical fiber (Thorlabs FGA105-LCA) after etching away the 20μm silica cladding from a 3cm section in the center of a 2.5m segment with buffered hydrofluoric acid (hazardous). The ITO film was deposited onto the exposed fiber core from an alkoxide solution, containing $InCl_3$, $Sn[OCH(CH_3)_2]_4$, $C_5H_8O_2$, and $C_3H_8O_2$ with a molar ratio of 1:0.065:5.9:28, by dragging a generated droplet at the tip of a 50μL micro pipette upwards over the exposed core, after which it was calcined in 1%$H_2$ at 600°C inside of a controlled environment tube furnace. The ITO surface plasmon resonance was characterized both by probing with an external light source (Ocean Optics DH-2000-bal), and by observing the thermally emitted light contained in the fiber with an NIR spectrometer (ArcOptics FTNIR-U-09-026). FIB lift-out was conducted with an FEI Nova Nanolab 600 dual beam FIB with a gallium beam, after coating the surface with platinum to maintain film integrity. Cross-sectional imaging and spot scan EDS analysis was conducted with a 200kV FEI Tecnai F20 TEM. The back-scattered electron image was obtained with an FEI Quanta 600F SEM.

## Acknowledgements

This work was funded by the U.S. DOE Advanced Research/Crosscutting Technologies program at the National Energy Technology Laboratory. This research was supported in part by an appointment to the National Energy Technology Laboratory Research Participation Program, sponsored by the U.S. Department of Energy and administered by the Oak Ridge Institute for Science and Education.

This report was prepared as an account of work sponsored by an agency of the United States Government. Neither the United States Government nor any agency thereof, nor any of their employees, makes any warranty, express or implied, or assumes any legal liability or responsibility for the accuracy, completeness, or usefulness of any information, apparatus, product, or process disclosed, or represents that its use would not infringe privately owned rights. Reference herein to any specific commercial product, process, or service by trade name, trademark, manufacturer, or otherwise does not necessarily constitute or imply its endorsement, recommendation, or favoring by the United States Government or any agency thereof. The views and opinions of authors expressed herein do not necessarily state or reflect those of the United States Government or any agency thereof.